\documentclass{emulateapj}

\usepackage{lineno}
\usepackage{amsmath}
\usepackage{natbib}
\usepackage{amssymb}
\usepackage{epstopdf}
\usepackage{graphicx}
\usepackage{subfigure} 
\usepackage{dcolumn}
\usepackage{bm}
\usepackage{rotating}
\usepackage{hyperref}

\shorttitle{Observation of an Anisotropy in the Galactic Cosmic Ray arrival direction at 400 TeV with IceCube}
\shortauthors{IceCube Collaboration: Abbasi et al.}

\begin{document}

\title{Observation of Anisotropy in the Galactic Cosmic Ray arrival directions at 400 TeV with IceCube}

\author{
IceCube Collaboration:
R.~Abbasi\altaffilmark{1},
Y.~Abdou\altaffilmark{2},
T.~Abu-Zayyad\altaffilmark{3},
M.~Ackermann\altaffilmark{4},
J.~Adams\altaffilmark{5},
J.~A.~Aguilar\altaffilmark{1},
M.~Ahlers\altaffilmark{6},
M.~M.~Allen\altaffilmark{7},
D.~Altmann\altaffilmark{8},
K.~Andeen\altaffilmark{1,9}
% ...
}
\altaffiltext{1}{Dept.~of Physics, University of Wisconsin, Madison, WI 53706, USA}
\altaffiltext{2}{Dept.~of Physics and Astronomy, University of Gent, B-9000 Gent, Belgium}
\altaffiltext{3}{Dept.~of Physics, University of Wisconsin, River Falls, WI 54022, USA}
\altaffiltext{4}{DESY, D-15735 Zeuthen, Germany}
\altaffiltext{5}{Dept.~of Physics and Astronomy, University of Canterbury, Private Bag 4800, Christchurch, New Zealand}
\altaffiltext{6}{Dept.~of Physics, University of Oxford, 1 Keble Road, Oxford OX1 3NP, UK}
\altaffiltext{7}{Dept.~of Physics, Pennsylvania State University, University Park, PA 16802, USA}
\altaffiltext{8}{III. Physikalisches Institut, RWTH Aachen University, D-52056 Aachen, Germany}
\altaffiltext{9}{now at Dept. of Physics and Astronomy, Rutgers University, Piscataway, NJ 08854, USA}
% ...

\begin{abstract}
				   
In this paper we report the first observation in the Southern hemisphere of an energy dependence in the Galactic cosmic ray anisotropy up to a few hundred TeV. This measurement was performed using cosmic ray induced muons recorded by the partially deployed IceCube observatory between May 2009 and May 2010. The data include a total of 33$\times 10^{9}$ muon events with a median angular resolution of $\sim3^{\circ}$ degrees. A sky map of the relative intensity in arrival direction over the Southern celestial sky is presented for cosmic ray median energies of 20 and 400 TeV. The same large-scale anisotropy observed at median energies around 20 TeV is not present at 400 TeV. Instead, the high energy skymap shows a different anisotropy structure including a deficit with a post-trial significance of \mbox{-6.3}$\sigma$. This anisotropy reveals a new feature of the Galactic cosmic ray distribution, which must be incorporated into theories of the origin and propagation of cosmic rays. 

\end{abstract}

\section{Introduction}
\label{sec:intro}

During the last decades, Galactic cosmic rays have been found to have a small but measurable energy dependent sidereal anisotropy in their arrival direction distribution with a relative amplitude of order of $10^{-4}$ to $10^{-3}$.
The first comprehensive observation of the cosmic ray sidereal anisotropy was provided by a network of muon detectors sensitive to cosmic rays between 10 and several hundred GeV~\citep{nagashima}. More recent underground and surface array experiments in the Northern hemisphere have shown that a sidereal anisotropy is present in the TeV energy range (Tibet Air Shower gamma (AS$\gamma$) array~\citep{amenomori}, Super-Kamiokande~\citep{guillian}, Milagro~\citep{abdo}, and ARGO-YBJ~\citep{zhang}). Furthermore, the IceCube neutrino observatory reported the first observation of a cosmic ray anisotropy in the Southern sky at energies in excess of about 10 TeV~\citep{abbasi}. The cosmic ray anisotropies reported by IceCube showed that the large scale features appeared to be a continuation of those observed in the Northern hemisphere.

At high energies, the Tibet AS$\gamma$ collaboration reported an observation for primary energies $\sim$300 TeV to be consistent with cosmic ray isotropic intensity~\citep{amenomori}, while the EAS-TOP collaboration reported a sharp increase in the anisotropy for primary energies $\sim$370 TeV~\citep{aglietta}. At the time of the writing of this paper the observations in the Northern hemisphere do not provide a coherent global picture of the sidereal anisotropy at high energy.  

The origin of the anisotropic distribution in the arrival direction of Galactic cosmic rays over the entire celestial sky is still unknown. If there is a relative motion of the solar system with respect to the cosmic ray plasma, then this would produce a well defined anisotropy. For example, if cosmic rays are at rest with respect to the galactic center, a dipole anisotropy would be expected. The magnitude of the anisotropy is calculated to be $0.35\%$ with an apparent excess of cosmic ray counts toward the direction of solar Galactic rotation ($\alpha =$ 315$^{\circ}$,$\delta =$48$^{\circ}$) and a deficit in the opposite direction ($\alpha =$ 135$^{\circ}$,$\delta =$\mbox{-48}$^{\circ}$). Such a dipole anisotropy is referred to as the Compton-Getting effect~\citep{comptongetting}. Neither the amplitude nor the phase expected from the Compton-Getting effect are consistent with the cosmic ray anisotropy observations (IceCube~\citep{abbasi}, Tibet Air Shower gamma (AS$\gamma$) array~\citep{amenomori}, Milagro~\citep{abdo}). Moreover, the observed sidereal anisotropy is not consistent with a simple dipole~\citep{abbasi}. It is worth noting that since the reference frame of the Galactic cosmic rays is not known, it is reasonable to assume that the Compton-Getting effect could be (at most) one of several contributions to the cosmic ray anisotropy.

While the origin of the anisotropy is not understood, it has been speculated that it might be a natural consequence of the distribution of cosmic ray Galactic sources, in particular nearby and recent supernova remnants (SNR). The discreteness of such sources, along with cosmic ray propagation through a highly heterogeneous interstellar medium, might lead to significant fluctuations of their intensity in space and time and, therefore, to an anisotropy in the arrival direction of cosmic rays at Earth~\citep{erlykin}. This speculation is challenged by~\cite{butt}, who points out that the observed anisotropy is of low intensity, whereas the high energy cosmic rays from such sources would escape the galaxy relatively quickly, leading to high anisotropy.

The study of the cosmic ray arrival distribution might provide hints into the properties of cosmic ray propagation in the turbulent interstellar magnetic field~\citep{beresnyak}. While at TeV energies it is speculated that propagation effects could either generate large scale anomalies in their arrival direction~\citep{battaner} or produce localized excess regions~\citep{malkov}, depending on the turbulence scale and diffusion properties, it is still not clear whether such models would be able to explain the observations at higher energies.

In this paper we present the analysis of cosmic ray data collected by the IceCube observatory, which we use to extend the observations of the Galactic cosmic ray anisotropies by IceCube ~\citep{abbasi},~\citep{abbasi11} up to several hundred TeV. The analysis procedure is described in Section~\ref{sec:analysis} and the anisotropy in sidereal reference frame is shown in Section~\ref{sec:results}. Section~\ref{sec:results} describes an experimental procedure to verify that the observed sidereal anisotropy is not an artifact of the analysis procedure, using the arrival distribution of cosmic rays as a function of the angular distance from the Sun. In this coordinate system, a dipole effect is expected such that the cosmic ray count rate is higher toward the direction of Earth's motion around the Sun and lower in the opposite direction. The experimental systematic uncertainties on the anisotropy in sidereal coordinates are described in Section~\ref{sec:syst} and the conclusions are summarized in Section~\ref{sec:conc}.

\section{Analysis}
\label{sec:analysis}

\subsection{Data and Reconstruction}
\label{sec:data}

IceCube is a neutrino observatory located at the geographic South Pole. During the 2009-2010 austral summer, the partially deployed detector was equipped with 3,540 Digital Optical Modules (DOMs) buried between about 1.5 and 2.5 km below the surface of the ice along 59 vertical strings~\citep{dommb}. The IceCube physics runs in the 59-string configuration (IceCube-59) started on May 20, 2009, and ended on May 30, 2010. IceCube observes relativistic charged particles by detecting the Cherenkov light produced as they travel through the ice. In particular the observatory is sensitive to the charged particles produced by neutrino interactions inside the ice, as well as the muons created in the cosmic ray air showers. 

In order to reject the random signals derived from the $\sim$500 Hz dark noise rate from each DOM, a local coincidence was required between neighboring DOMs with a coincidence time interval of $\pm$ 1,000 ns. A trigger was then produced when eight or more DOMs in local coincidence detected photons within 5,000 ns. The trigger rate in IceCube-59, predominantly from muons produced in cosmic ray air showers, ranged from a minimum of about 1,600 Hz in the austral winter to a maximum of about 1,900 Hz in the austral summer. This modulation is due to the large seasonal variation of the stratospheric temperature, and consequently the density, which affects the decay rate of mesons into muons~\citep{tilav}. 

All recorded events were processed using a coarse online fit to their trajectories~\citep{muonreco}. To refine the directional estimate, the coarse fit was used to seed an online likelihood-based reconstruction, which was applied if ten or more optical sensors were triggered by the event. The average rate of the events that passed the likelihood-based reconstruction ranged from a minimum of about 1,150 Hz to a maximum of about 1,350 Hz. All the events collected and processed by the IceCube observatory were stored in a compact Data Storage and Transfer format, or DST, and shipped North through satellite link (see \cite{abbasi11} for details). This analysis uses all events with likelihood directional reconstruction stored in the DST data format, collected within one full calendar year from the beginning of the run on May 20, 2009. After rejecting short data runs we ended up with $33 \times 10^{9}$ events, corresponding to a detector livetime of 324.8 days. The events have a median angular resolution of about $3^{\circ}$ and a median energy of the cosmic ray parent particles of about 20 TeV. It is worth noting that this angular resolution is a property of this data sample and the applied reconstruction algorithms; reduced data samples using more advanced reconstructions for high energy neutrino searches have a typical angular error less than $1^{\circ}$~\citep{pointsourcepaper}.

\begin{figure}[!ht]
\subfigure[] 
{
   \includegraphics[width=0.5\textwidth]{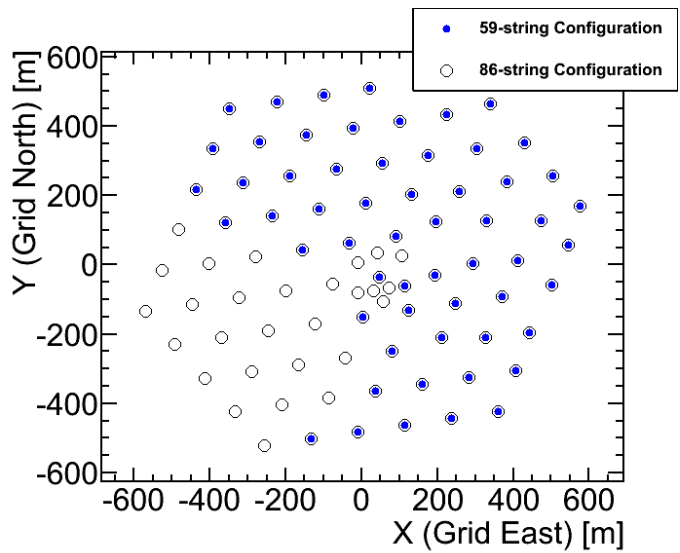}
}
\hspace{1cm}
 \subfigure[] 
{
   \includegraphics[width=0.5\textwidth]{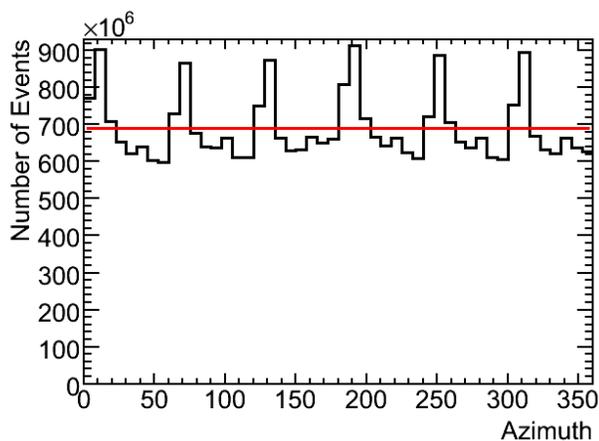}
}
\hspace{1cm}

\caption{\label{AzimuthDist}
Figure (a) shows the complete IceCube 86-string configuration. Circles filled in blue represent the IceCube 59-string configuration which is the main configuration used in this paper.
Figure (b) shows the azimuth distribution for the whole data set. It shows the number of events vs. the azimuth of the arrival direction of the primary cosmic ray particle. The horizontal red line is the average number of events for the distribution.
}
\end{figure}

To measure an anisotropy of order $10^{-4}$ to $10^{-3}$, it is necessary to eliminate any background effects that could mimic such an observation. Due to its unique location at the geographic South Pole, the IceCube observatory has full coverage of the Southern sky at any time of the year. Therefore, seasonal variations in the muon intensity occur uniformly across the entire field of view and do not affect the local arrival direction distribution of the reconstructed events~\citep{abbasi}. The main effect that needs to be accounted for is due to the geometrical shape of IceCube: the hexagonal geometric structure of the observatory introduces a strong asymmetry in the local azimuth distribution of events (Figure~\ref{AzimuthDist}). Non-uniform time coverage caused by detector downtime and run selection reduces the total detector livetime by about $\sim$ 10\%, preventing the complete averaging of the local coordinate asymmetry over one year and generating spurious variations in the arrival directions of cosmic rays in celestial coordinates. To remove this effect, the asymmetry in the local azimuthal acceptance (shown in Figure~\ref{AzimuthDist}-b) is corrected by re-weighting the number of events from a local azimuth bin to the average number of events over the full range of the local azimuth distribution. This re-weighting is applied in four zenith bands with approximately the same number of events per band due to the detector azimuth distribution variation with zenith angle~\citep{abbasi}.

\begin{figure}[!ht]
\includegraphics[width=0.5\textwidth]{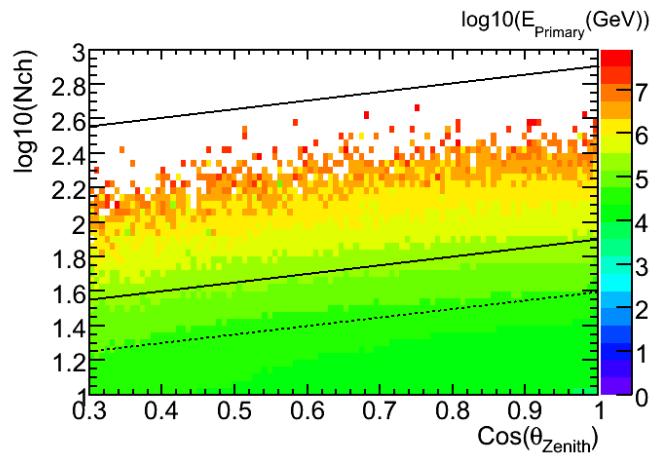}
\caption{\label{EReco}
The average logarithm of the cosmic ray primary energy as a function of $N_\text{ch}$ and Zenith angle, as obtained from simulation. The Y-axis is the log$_{10}$ of $N_\text{ch}$, the X-axis is the cosine of the reconstructed Zenith angle of the event while the color scale is the mean of the logarithm of the cosmic ray primary energy for each bin obtained from simulation in GeV. The first energy band with median energy of 20 TeV is all the events selected below the dashed line, while the second energy band with median energy of 400 TeV contains events selected between the continuous lines.}

\end{figure}

\subsection{Estimation of Cosmic Ray Energy}
\label{ssec:energy}

Since IceCube detects cosmic rays indirectly through the observation of muons produced in extensive air showers, the energy of the cosmic ray primary particle is estimated based on the total amount of light seen by the detector, which is  a function of the number and energy of detected muons. Muons produced in the atmosphere propagate through the ice losing energy via ionization and stochastic processes such as pair production, bremsstrahlung and photo-nuclear interactions. The secondary charged particles produced by these processes emit Cherenkov light. The number of emitted photons is proportional to the total energy of the secondaries. By detecting photons, it is possible to estimate the energy lost by the muons and therefore the muon's energy within the volume instrumented with optical sensors. However, the total energy of the detected muons is only a fraction of the original cosmic ray primary energy, while the rest is mostly dispersed into the electromagnetic component of the air shower. As a consequence, the natural fluctuations that arise in the development of the extensive air showers limit the resolution of the estimate of the primary energy that one can make using muons in ice.

The uncertainty in the cosmic ray energy estimation has been modeled with a full simulation of cosmic ray interactions in the atmosphere using CORSIKA~\citep{sim1} with SIBYLL hadronic interaction model (Version 2.1)~\citep{sim2} together with the composition and the spectrum of primary cosmic rays as described in~\cite{sim3}. Muons were propagated through the ice with the Muon Monte Carlo (MMC) propagator~\citep{mmc}, and a full detector simulation was performed on those events. 

\begin{figure}[!ht]
\includegraphics[width=0.5\textwidth]{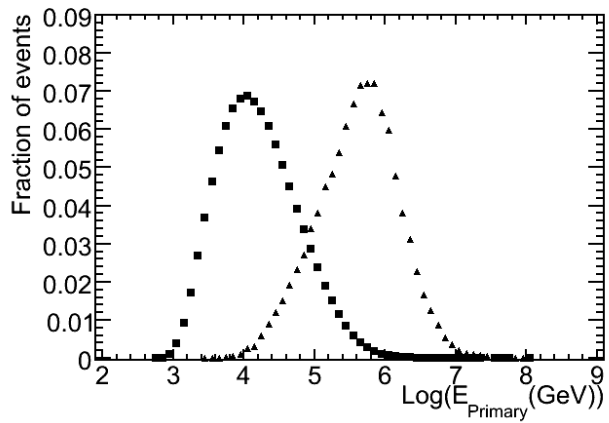}
\caption{\label{FracvsErecoNch}
The fraction of events vs. the logarithm of primary energy (in GeV) for the two selected energy samples (see text). The low energy sample contains events with a median energy of 20 TeV (squares) and the high energy sample contains events with a median energy of 400 TeV (triangles). The energy distributions were determined using a full  simulation of cosmic ray interactions in the atmosphere, of muons propagation through the ice and of the IceCube-59 detector.}
\end{figure}
\begin{figure}[!ht]
\includegraphics[width=0.5\textwidth]{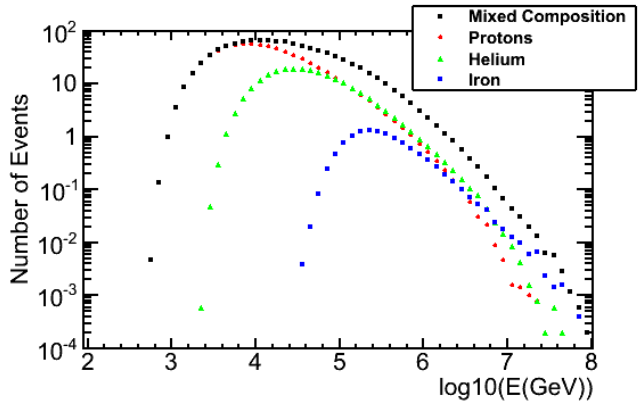}
\caption{\label{spec}
The number of events seen by IceCube vs. the logarithm of primary energy (in GeV) using the composition model described in~\cite{sim3}. Fractional contributions of proton, helium, and iron are shown as well. At 20 TeV, the spectrum is dominated by the proton fractional contribution of $\sim 70\%$, while at 400 TeV that fraction will have decreased to $\sim 30\%$. The energy distributions were determined using a full simulation of cosmic ray interactions in the atmosphere as described in this section.}
\end{figure}

In this analysis the estimate of the cosmic ray energy is based on the number of DOMs hit by Cherenkov photons (i.e. number of channels, or $N_\text{ch}$). The downward muons reaching IceCube with a large zenith angle $\theta$ have to cross a larger slant depth than vertically propagating muons, and so the set of horizontal events naturally excludes lower primary energy cosmic rays. This introduces a zenith angle dependence of the relation between $N_\text{ch}$ and the primary particle energy. Therefore, a two-dimensional cut in $N_\text{ch}$ and $\theta$ is used. Figure~\ref{EReco} shows the distribution from simulation of the cosmic ray primary particle energy with respect to $N_\text{ch}$ as a function of $\cos \theta$. The figure shows that for a given range of $N_\text{ch}$, vertical events (i.e. $\cos \theta$ $\approx$ 1) are dominated by cosmic rays with lower average energy than horizontal events (i.e. $\cos \theta$ $\approx$ 0.3) due to the larger ice thickness the muons would go through before triggering the detector. We identified regions of constant primary energy in ($N_\text{ch}$, $\cos \theta$), delimited with the black lines in Figure~\ref{EReco}, in order to select two event samples at energies with minimal overlap and, at the same time, with the maximum possible number of events in the high energy sample. The low energy sample was obtained by selecting all events below the dashed line in Figure~\ref{EReco}, and the high energy sample by selecting events between the solid lines in the figure.

Figure~\ref{FracvsErecoNch} shows the simulated primary energy distributions for the two event samples. The estimate of the primary cosmic ray energy has a resolution of about 0.5 in the logarithmic scale. The uncertainty of the primary energy estimate is dominated by the fluctuations in the air showers. The low energy sample over the Southern sky contains $21\times10^9$ events; assuming the composition described by~\cite{sim3} and shown in Figure~\ref{spec}. The median primary particle energy of the low energy sample is 20 TeV, with $68\%$ of the events are between $4-63$ TeV.  The high energy sample contains $0.58\times10^9$ events. The median primary particle energy of the high energy sample is 400 TeV, with $68\%$ of the events are between $100-1,258$ TeV.

\section{Results}
\label{sec:results}

\subsection{Sidereal Anisotropy}
\label{ssec:sidereal}

In order to investigate the cosmic ray arrival direction distribution, we determine the map of deviation from isotropy by calculating the relative intensity distribution after azimuthal re-weighting of the arrival directions of the data as described in the previous section. The cosmic ray arrival direction distribution is dominated by the zenith angle dependence of the muon flux. The zenith angle dependence is a result of a varying overburden for the muons through the ice. Therefore, the flux for each bin is normalized within each zenith band (or, equivalently at the South Pole, each declination band):

\begin{equation}
I_i = \frac{N_i(\alpha, \delta)}{\langle N_i(\delta)\rangle_{\alpha}},
\label{relint}
\end{equation}
where $I_i$  is the relative intensity for each bin of angular equatorial coordinates ($\alpha$, $\delta$), $N_i$ is the number of events in bin i, and $\langle N_i\rangle$ is the average number of events for the bins along the same iso-latitude as bin i (with the same declination $\delta$). The sky maps in this analysis are produced using the Hierarchical Equal Area IsoLatitude Pixelization (HEALPix) libraries~\citep{healpix}. HEALPix subdivides the unit sphere into quadrilateral pixels of equal area. In this analysis, the maps contain pixels that correspond to an angular resolution of $\sim 3^{\circ}$, which approximately corresponds to the angular resolution of the detector.

\begin{figure}[!ht]

 \subfigure[] 
{
      \includegraphics[width=0.5\textwidth]{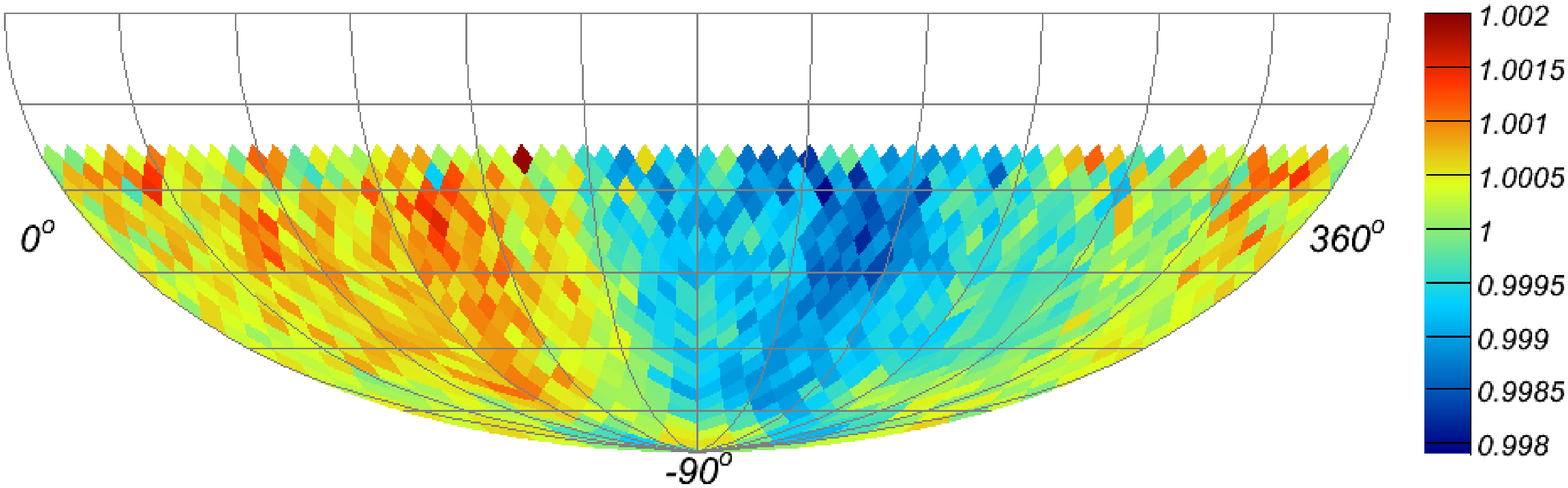}
     
}
\hspace{1cm}
 \subfigure[] 
{
      \includegraphics[width=0.5\textwidth]{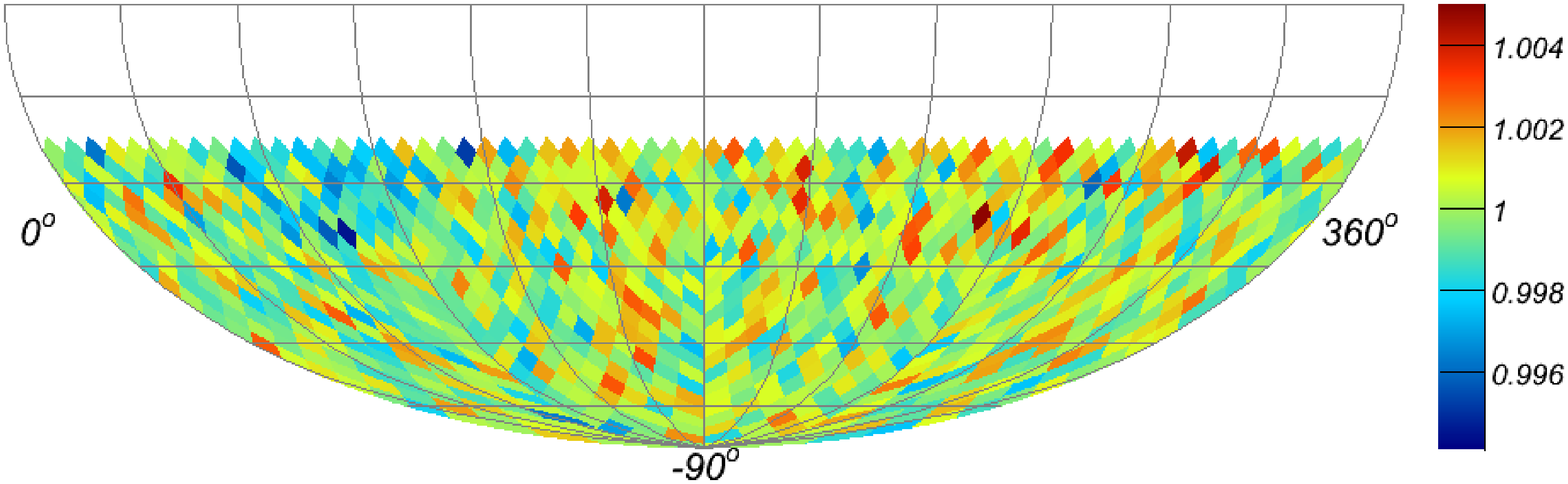}
}
\hspace{1cm}

\caption{\label{sidskymap2d} Figure (a) shows the IceCube cosmic ray map of the first energy band (median energy of 20 TeV) for the relative intensity in right ascension $\alpha$. Figure (b) shows the IceCube cosmic ray map for the second energy band (median energy of 400 TeV) of the relative intensity in right ascension $\alpha$.
}

\end{figure}

Figure~\ref{sidskymap2d} show the maps of the relative intensity in cosmic ray arrival direction in sidereal reference frame (equatorial coordinates), for the low and high energy samples, respectively. 
The color scale in the figures represents the relative intensity as described in eq.~\ref{relint}. The observed sidereal anisotropy appears to evolve as a function of energy and the anisotropy pattern observed at 400 TeV shows substantial differences with respect to that observed at 20 TeV. Note that in the maps only the pixels below declination angle of \mbox{-25$^{\circ}$} are shown. Pixels above declination of \mbox{-25$^{\circ}$} are masked due to the degradation of the angular resolution at higher declinations. Such degradation is to be expected because of the poorer statistical power and the domination by mis-reconstructed events ~\citep{abbasi11}.

\begin{figure}[!ht]
\subfigure[] 
{
      \includegraphics[width=0.5\textwidth]{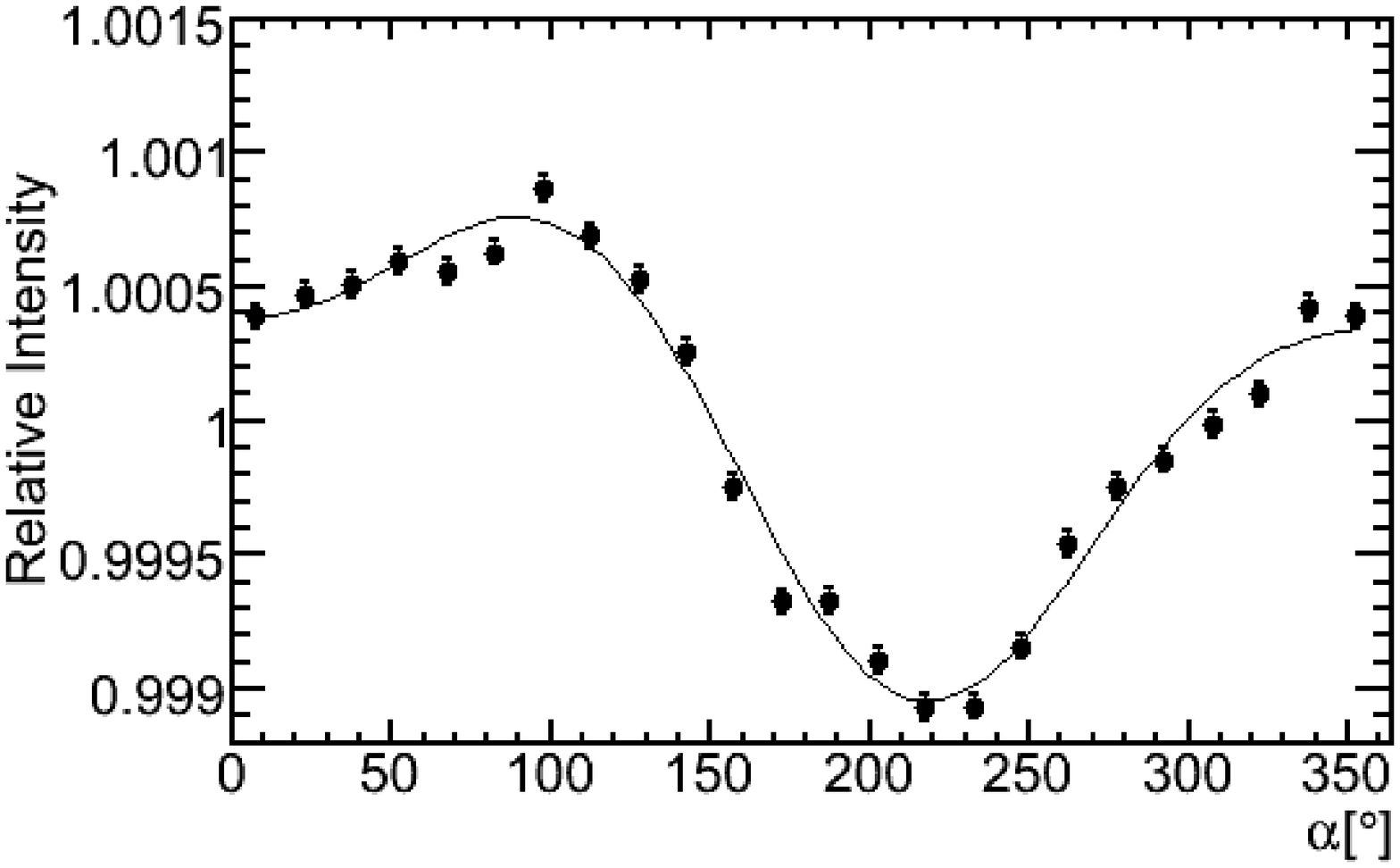}
     
}
 \subfigure[] 
{
      \includegraphics[width=0.5\textwidth]{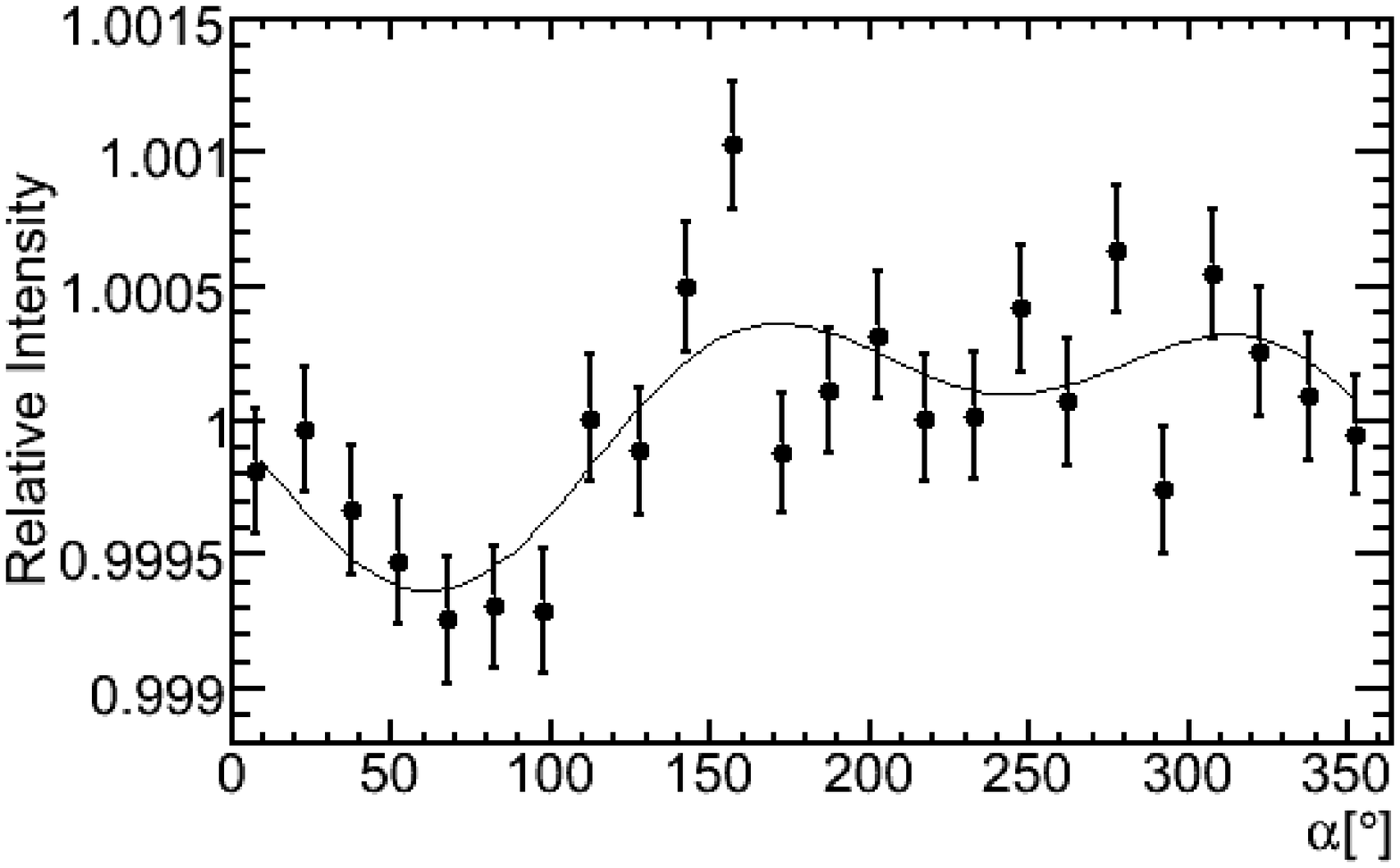}
}
\hspace{1cm}
\caption{\label{sidskymap1d}
Figure (a) shows the one-dimensional projection in right ascension $\alpha$ of the first energy band (20 TeV) of two-dimensional cosmic ray map in Figure~\ref{sidskymap2d}-a. Figure (b) shows the  one-dimensional projection in right ascension $\alpha$ of the second energy band (400 TeV) of two-dimensional cosmic ray map in Figure~\ref{sidskymap2d}-b. The data are shown with statistical uncertainties, and the black line corresponds to the first and second harmonic fit to the data. 
}
\end{figure}

In order to characterize quantitatively the general structure of the anisotropy, we proceed as follows.  For each row of pixels in the map, a 24-bin histogram is made from the relative intensity values of the pixels (where each pixel's value is included in the bin which contains the right ascension of the center of the pixel).  The rows are spaced approximately every $\sim$3 degrees in declination, and the histograms are constructed down to declination \mbox{-72} degrees (beyond which the number of pixels per declination band is less than the number of bins in the histogram).  The binned relative intensity data were then fitted to a harmonic function of the form
\begin{equation}
\sum_{j=1}^{2}A_{j} \cos[j(\alpha-\phi_{j})] + B, 
\label{eq1}
\end{equation}
where $j$ is the harmonic term order (i.e. dipole for $j$=1, quadrupole for $j$=2), $A_{j}$ is the amplitude of the $j^{th}$ harmonic term, $\phi_{j}$ is the phase of the $j^{th}$ harmonic term, $\alpha$ is the right ascension, and $B$ is a constant. The results of this fit are shown in Table~\ref{tab1} and Table~\ref{tab2} for the low and high energy samples, respectively. In addition, in order to quantify the sidereal anisotropy over the whole Southern hemisphere, the anisotropy profile in right ascension is measured by accumulating the relative intensity distribution from the declination belts. The error bars were obtained by propagating the statistical errors from each declination belt. Figure~\ref{sidskymap1d} show the projections in right ascension of the cosmic ray relative intensity in sidereal reference frame, for the low and high energy samples, respectively. The lines in the figures represent the fit to the first and second harmonic terms of eq. \ref{eq1}, and the fit results are shown in Table~\ref{sidvals} together with the $\chi^2$/ndof values for the first and second harmonic fits, in addition to the number of events used in the right ascension projections. While the $\chi^2$/ndof indicates that the fits do not completely describe the data, we found that even fitting up to the sixth harmonic does not completely fit all of the structures, so we use here only the first and second harmonics as a general characterization of the anisotropy.


\begin{table}[!ht]
\begin{ruledtabular}  
\caption{\label{tab1}Harmonic fit values per declination band for the energy band centered at 20 TeV.}

\begin{tabular}{cllll}
Dec. &  $A_{1}$ $\pm$ (stat.) & $\phi_{1} \pm$ (stat.) &  $A_{2}$ $\pm$ (stat.) & $\phi_{2} \pm$ (stat.)\\
Mean &  $10^{-4}$& [$^{\circ}$] & $10^{-4}$  & [$^{\circ}$] \\
\hline

-24 & 7.1 $\pm$ 1.0 & 37.3 $\pm$ 8.1 & 3.2 $\pm$ 1.0 & 303.5 $\pm$ 9.0\\
-27 & 8.4 $\pm$ 0.9 & 35.6 $\pm$ 6.0 & 2.1 $\pm$ 0.9 & 321.3 $\pm$ 11.8\\
-30 & 8.7 $\pm$ 0.7& 45.4 $\pm$ 4.7 & 4.0 $\pm$ 0.7 & 306.6 $\pm$ 5.1\\
-33 & 8.6 $\pm$ 0.7 & 50.5 $\pm$ 4.3 & 3.6 $\pm$ 0.7 & 294.6 $\pm$ 5.0\\
-36 & 9.3 $\pm$ 0.5 & 51.2 $\pm$ 3.3 & 3.1 $\pm$ 0.5 & 299.1 $\pm$ 5.0\\
-39 & 8.3 $\pm$ 0.5 & 52.9 $\pm$ 3.4 & 2.1 $\pm$ 0.5 & 299.6 $\pm$ 6.6\\
-42 & 9.6 $\pm$ 0.4 & 51.1 $\pm$ 2.6 & 3.1 $\pm$ 0.4 & 301.8 $\pm$ 4.0\\
-45 & 9.3 $\pm$ 0.4 & 57.4 $\pm$ 2.8 & 3.0 $\pm$ 0.5 & 305.9 $\pm$ 4.2\\
-48 & 8.0 $\pm$ 0.4 & 56.7 $\pm$ 2.8 & 2.7 $\pm$ 0.4 & 304.3 $\pm$ 4.0\\
-51 & 7.9 $\pm$ 0.4 & 57.2 $\pm$ 2.8 & 2.5 $\pm$ 0.4 & 293.0 $\pm$ 4.3\\
-54 & 8.0 $\pm$ 0.4 & 55.9 $\pm$ 2.6 & 2.3 $\pm$ 0.4 & 297.9 $\pm$ 4.5\\
-57 & 7.9 $\pm$ 0.4 & 60.8 $\pm$ 2.7 & 1.8 $\pm$ 0.4 & 303.3 $\pm$ 5.6\\
-60 & 7.9 $\pm$ 0.4 & 52.7 $\pm$ 2.6 & 2.0 $\pm$ 0.4 & 300.4 $\pm$ 5.3\\
-63 & 7.7 $\pm$ 0.4 & 49.9 $\pm$ 3.3 & 1.8 $\pm$ 0.4 & 307.1 $\pm$ 6.7\\
-66 & 7.3 $\pm$ 0.4 & 51.0 $\pm$ 2.9 & 4.1 $\pm$ 0.4 & 293.2 $\pm$ 2.7\\
-69 & 5.7 $\pm$ 0.4 & 50.8 $\pm$ 4.2 & 4.9 $\pm$ 0.4 & 282.4 $\pm$ 2.4\\
-72 & 5.7 $\pm$ 0.4 & 38.8 $\pm$ 4.0 & 3.6 $\pm$ 0.4 & 301.7 $\pm$ 3.2\\

\end{tabular}

\end{ruledtabular}   
\tablecomments{First and second harmonic fit values per declination for the first energy band.}
\end{table}


\begin{table}[!ht]
\begin{ruledtabular}  
\caption{\label{tab2} Harmonic fit values per declination band for the energy band centered at  400 TeV.}

\begin{tabular}{clllll}
Dec. & $A_{1}$ $\pm$ (stat.)& $\phi_{1} \pm$ (stat.) &
$A_{2}$ $\pm$ (stat.)& $\phi_{2} \pm$ (stat.)\\
Mean & $10^{-4}$ & [$^{\circ}$]& $10^{-4}$ &[$^{\circ}$] \\
\hline

-24 & 9.6 $\pm$ 3.1 & 248.1 $\pm$ 18.6 & 5.4 $\pm$ 3.1 & 143.6 $\pm$ 16.6 \\
-27 & 1.1 $\pm$ 3.0 & 245.7 $\pm$ 15.8 & 6.5 $\pm$ 3.0 & 158.1 $\pm$ 13.2 \\
-30 & 5.1 $\pm$ 2.6 & 238.9 $\pm$ 29.6 & 3.0 $\pm$ 2.6 & 146.9 $\pm$ 25.2 \\
-33 & 3.9 $\pm$ 2.7 & 255.9 $\pm$ 37.8 &   2.0 $\pm$ 2.6 &  205.3 $\pm$ 37.6 \\
-36 & 9.6 $\pm$ 2.4 & 217.0 $\pm$ 14.2 & 6.2 $\pm$ 2.4 & 171.5 $\pm$ 10.9 \\
-39 & 9.5 $\pm$ 2.4 & 246.9 $\pm$ 14.3 & 6.5 $\pm$ 2.4 & 144.2 $\pm$ 10.5 \\
-39 & 9.5 $\pm$ 2.4 & 246.9 $\pm$ 14.3 & 6.5 $\pm$ 2.4 & 234.2 $\pm$ 10.5 \\
-42 & 4.2 $\pm$ 2.2 & 246.2 $\pm$ 30.1 & 2.5 $\pm$ 2.2 & 231.3 $\pm$ 25.4 \\
-45 & 1.2 $\pm$ 2.5 & 311.4 $\pm$ 115.6 & 2.8 $\pm$ 2.5 & 110.4 $\pm$ 25.1 \\
-48 & 1.4 $\pm$ 2.3 & 181.0 $\pm$ 95.6 & 3.6 $\pm$ 2.3 & 154.2 $\pm$ 18.2 \\
-51 & 3.7 $\pm$ 2.4 & 236.7 $\pm$ 38.2 & 2.0 $\pm$ 2.4 & 156.8 $\pm$ 35.6 \\
-54 & 5.5 $\pm$ 2.4 & 220.8 $\pm$ 25.8 & 1.5 $\pm$ 2.5 & 142.5 $\pm$ 46.8 \\
-57 & 1.4 $\pm$ 2.6 & 228.8 $\pm$ 112.1 & 3.7 $\pm$ 2.6 & 165.0 $\pm$ 21.9 \\
-60 & 3.9 $\pm$ 2.6 & 359.8 $\pm$ 38.5 & 7.4 $\pm$ 2.6 & 161.0 $\pm$ 10.2 \\
-63 & 2.6 $\pm$ 3.4 & 13.0  $\pm$ 72.8 & 3.2 $\pm$ 3.3 & 148.6 $\pm$ 29.6 \\
-66 & 1.3 $\pm$ 2.9 & 143.4 $\pm$ 127.8 & 5.3 $\pm$ 3.0 & 107.5 $\pm$ 15.9 \\
-69 & 1.0 $\pm$ 3.4 & 304.5 $\pm$ 188.2 & 4.2 $\pm$ 3.4 & 227.9 $\pm$ 23.2 \\
-72 & 6.8 $\pm$ 3.4 & 174.8 $\pm$ 28.4 & 6.7 $\pm$ 3.4 & 152.5 $\pm$ 14.5 \\
\end{tabular}
\end{ruledtabular}   
\tablecomments{First and second harmonic fit values per declination for the second energy band. }
\end{table}

\begin{table*}[!ht]
\begin{ruledtabular}  
\begin{center}

\caption{\label{sidvals} In this table a summary of the sidereal anisotropy energy dependence is displayed. The first column is the median energy of the cosmic ray primary particles for the first and second energy band. The second column is the number of events used in the one-dimensional projection from declination -24 to declination -72. The values of the first and second harmonic fits amplitudes and phases together with their statistical and systematic uncertainties are displayed in column three through six. The last column is the $\chi^{2}/ndof$ for the first and second harmonic fit to the one-dimensional projection.}
\begin{tabular*}{\textwidth}{ccccccc}
\hline
$E_{Median}$ & events &  $A_{1_{SID}}$ & $\phi_{1_{SID}}$ & $A_{2_{SID}}$ & $\phi_{2_{SID}}$& $\chi^{2}/ndof$\\ 
(TeV)    &  ($10^{9}$)   & ($10^{-4}$) & (degree) & ($10^{-4}$) & (degree) & \\
\hline
20    & 17.9 &$7.9\pm0.1_{stat.}\pm0.3_{syst.}$ &$ 50.5\pm 1.0_{stat} \pm 1.1_{syst.} $ & $2.9 \pm 0.1_{stat.} \pm 0.4_{syst.}$ &  $ 299.5\pm 1.3_{stat} \pm 1.5_{syst.}$  & 95/19\\ \hline
400  & 0.5 & $3.7\pm0.7_{stat.}\pm0.7_{syst.}$ & $239.2 \pm 10.6_{stat} \pm 10.8_{syst.}$  &  $2.7 \pm 0.7_{stat.} \pm 0.6_{syst.}$ & $152.7 \pm 7.0_{stat} \pm 4.2_{syst.}$& 34/19\\ \hline

\hline
\end{tabular*}
\end{center}
\end{ruledtabular}   
\end{table*}


\subsubsection{Significance}

\begin{figure}[!ht]
\subfigure[] 
{
      \includegraphics[width=0.5\textwidth]{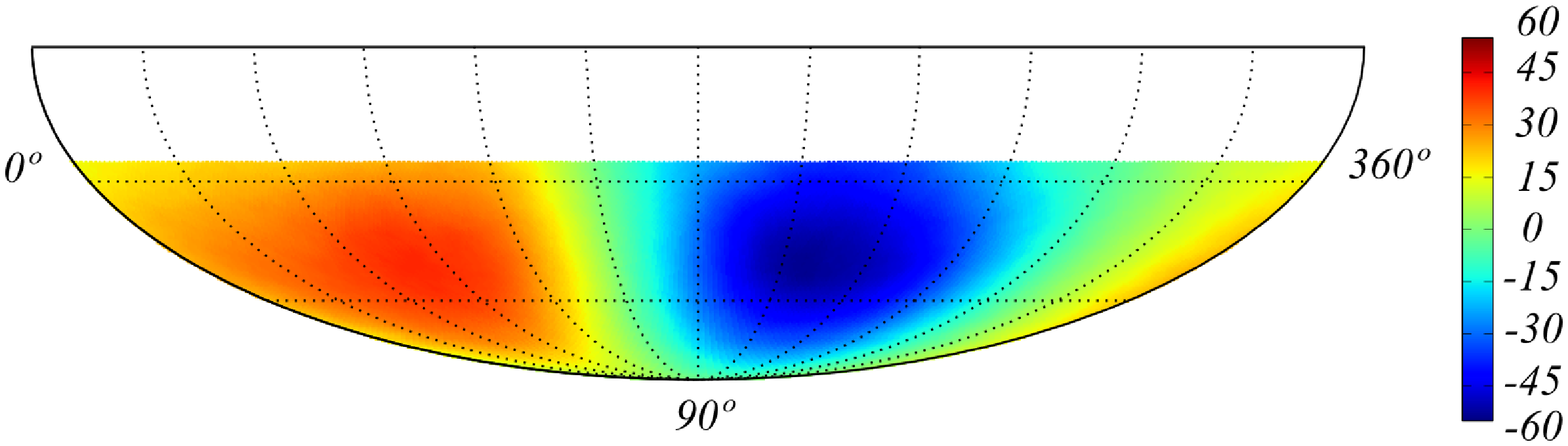}
     
}
 \subfigure[] 
{
      \includegraphics[width=0.5\textwidth]{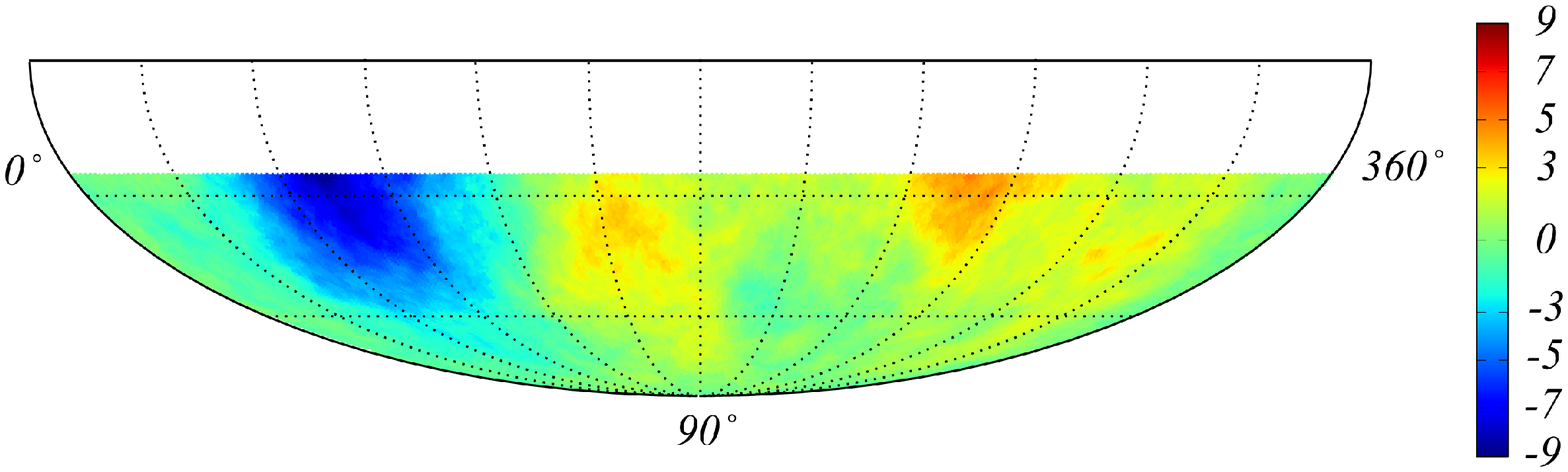}
}
\hspace{1cm}
\caption{\label{sigE1E2} Figure (a) shows the significance map for the 20 TeV energy band
 plotted with 30 degree smoothing.
Figure (b) shows the significance map for the 400 TeV energy band plotted with 20 degree smoothing.}
\end{figure}

Figure~\ref{sigE1E2}-a shows the significance map for the 20 TeV energy, while Figure~\ref{sigE1E2}-b shows the significance map for the 400 TeV energy. The significance skymaps are calculated using the direct integration method  with a time window of 24 hours and an optimized smoothing as described in~\cite{abbasi11}. The smoothing is then applied to the significance skymaps to improve the sensitivity to large features. The smoothing search applied in this analysis is from 1 to 30 degrees. After smoothing is optimized, the significance is then calculated using the method of~\cite{lima}.
 
The maximum significant features in the 20 TeV map with a 30 degree smoothing are found  with an excess at ($\alpha$ = 80.8$^{\circ}$, $\delta =$\mbox{-49.7}$^{\circ}$) with a significance value of 40$\sigma$, and a deficit at ($\alpha =$ 219.7$^{\circ}$,$\delta =$\mbox{-52.0}$^{\circ}$) with a significance value of \mbox{-53.5}$\sigma$. Moreover, for the 400 TeV map, two regions were identified to be significant. The first region is an excess at ($\alpha =$ 256.6$^{\circ}$,$\delta =$\mbox{-25.9}$^{\circ}$) with a significance of 5.3$\sigma$ and an optimized smoothing of 29 degrees, and the second region is a deficit at ($\alpha =$ 73.1$^{\circ}$,$\delta =$\mbox{-25.3}$^{\circ}$) with a significance of \mbox{-8.6$\sigma$} and an optimized smoothing of 21 degree. These significance values do not account for the scan for the peak significance in all pixels of the sky or the scan over smoothing radii applied to obtain an optimal sensitivity to the observed features.  We conservatively estimate a trial factor by assuming that all scans give statistically independent results.  After correcting for the trials, only the deficit remained significant beyond the \mbox{5}$\sigma$ level, with a  post-trial significance value of \mbox{-6.3}$\sigma$. This is the first significant observation of an anisotropy in the Southern sky at 400 TeV. The implications of this observation is explored in the conclusion and discussion sections of the paper.

\subsection{Solar Dipole Anisotropy}
\label{ssec:solar}

Currently there is no detailed theoretical model that predicts the observed sidereal anisotropy in the cosmic ray arrival direction distribution. Except for testing the stability of the Observatory and its time coverage (see sec.~\ref{sec:syst}), the only effective way to have an absolute calibration of the experimental sensitivity for the detection of the sidereal directional asymmetries is to measure the solar anisotropy from the Earth's orbital motion around the Sun. The solar anisotropy is well understood and was first reported in 1986 by~\cite{cutler} and then later observed by experiments in the multi-TeV energy range (Tibet AS$\gamma$~\citep{amenomori04, amenomori}, Milagro~\citep{abdo} and EAS-TOP~\citep{aglietta}). The observed solar anisotropy consists of a dipole that describes an apparent excess of cosmic rays in the direction of Earth's motion around the Sun and a deficit in the opposite direction. The relative intensity of the solar dipole is expressed as

\begin{equation}
\frac{\Delta I}{\langle I\rangle} = (\gamma +2)~\frac{v}{c}~\cos(\theta_v) ,
\label{eq:cg}
\end{equation}
where $I$ is the intensity, $\gamma$ the differential cosmic ray spectral index, $v$ the Earth's velocity, $c$ the speed of light, and $\theta_v$ the angle between the reconstructed arrival direction of the cosmic rays and the direction of motion of the observer~\citep{comptongetting, gleeson}. The actual amplitude of the observed solar dipole depends on the geographical latitude of the observer and on the angular distribution of the detected cosmic ray events at the observatory. 

Due to the location of IceCube at the South Pole, the sky is fully visible at any given time. Therefore, the solar anisotropy is observed in a reference system where the location of the Sun is fixed, where the latitude coordinate is the declination and the longitude is defined as right ascension difference of the cosmic ray arrival direction from the right ascension of the Sun ($\alpha - \alpha_{sun}$). In this reference frame the dipole excess is expected to be at 270$^{\circ}$ and the deficit at 90$^{\circ}$.

\begin{figure}[!ht]
 \subfigure[] 
{
      \includegraphics[width=0.5\textwidth]{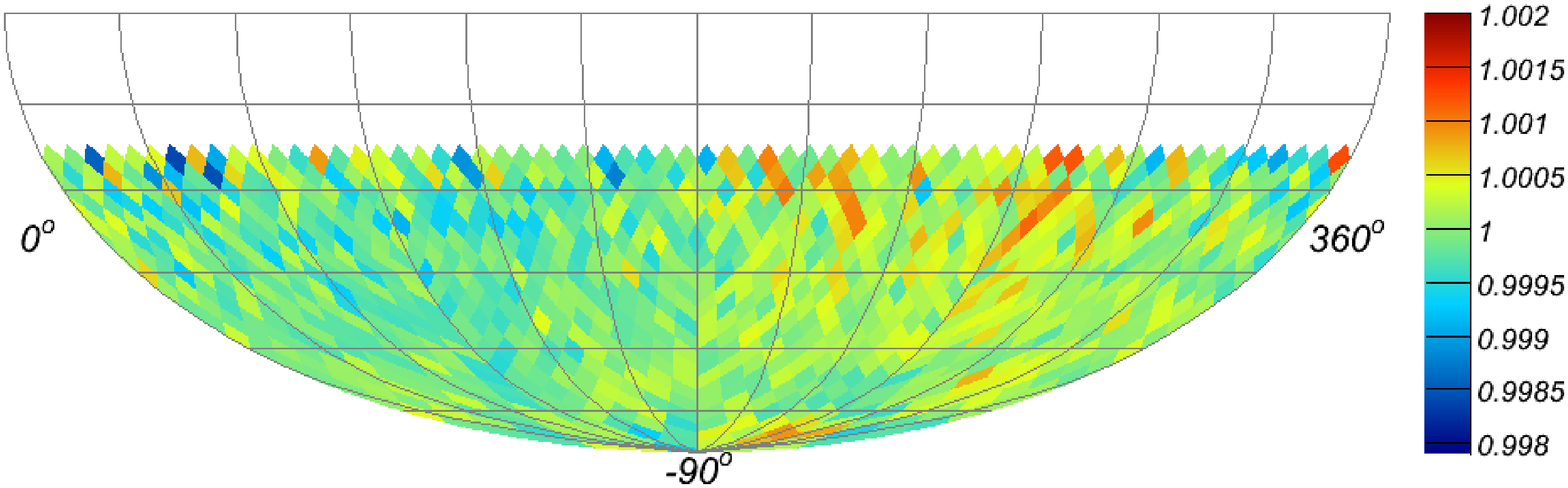}
     
}
\hspace{1cm}
\subfigure[] 
{
  \includegraphics[width=0.5\textwidth]{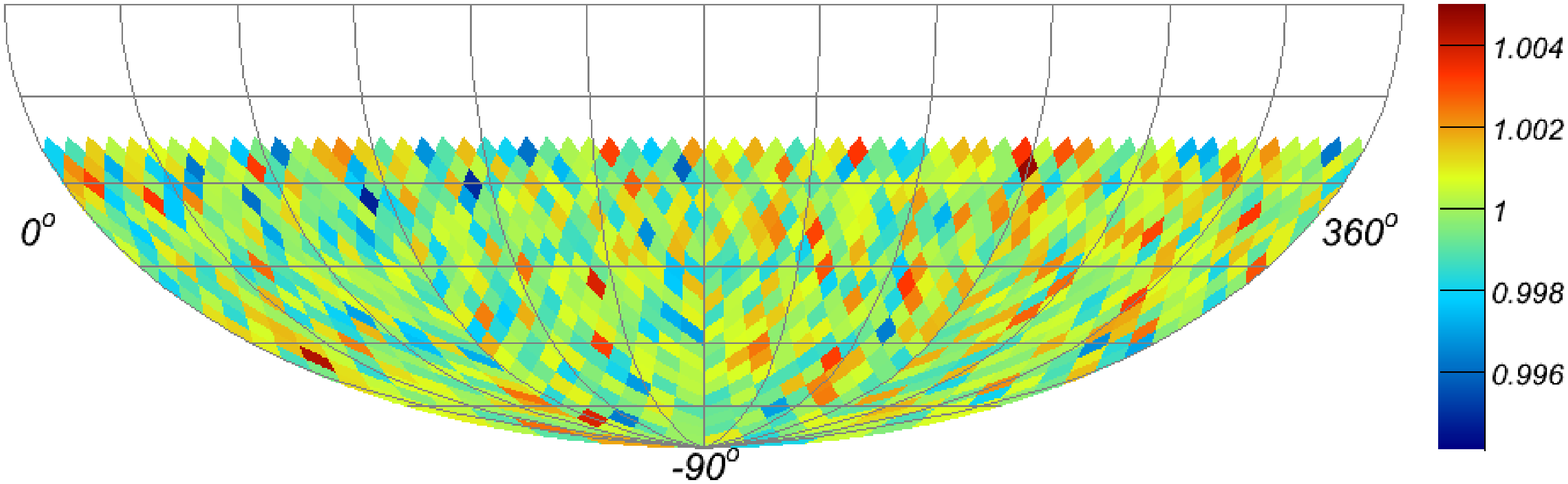} 
 
}
\hspace{1cm}
\caption{\label{solskymap2d} Figure (a) shows the IceCube cosmic ray map of the first energy band (median energy of 20 TeV) for the relative intensity in right ascension from the sun ($\alpha - \alpha_{sun}$). Figure (b) shows the IceCube cosmic ray map of the second energy band (median energy of 400 TeV) for the relative intensity in right ascension from the sun ($\alpha - \alpha_{sun}$).}

\end{figure}

\begin{figure}[!ht]

 \subfigure[] 
{
      \includegraphics[width=0.5\textwidth]{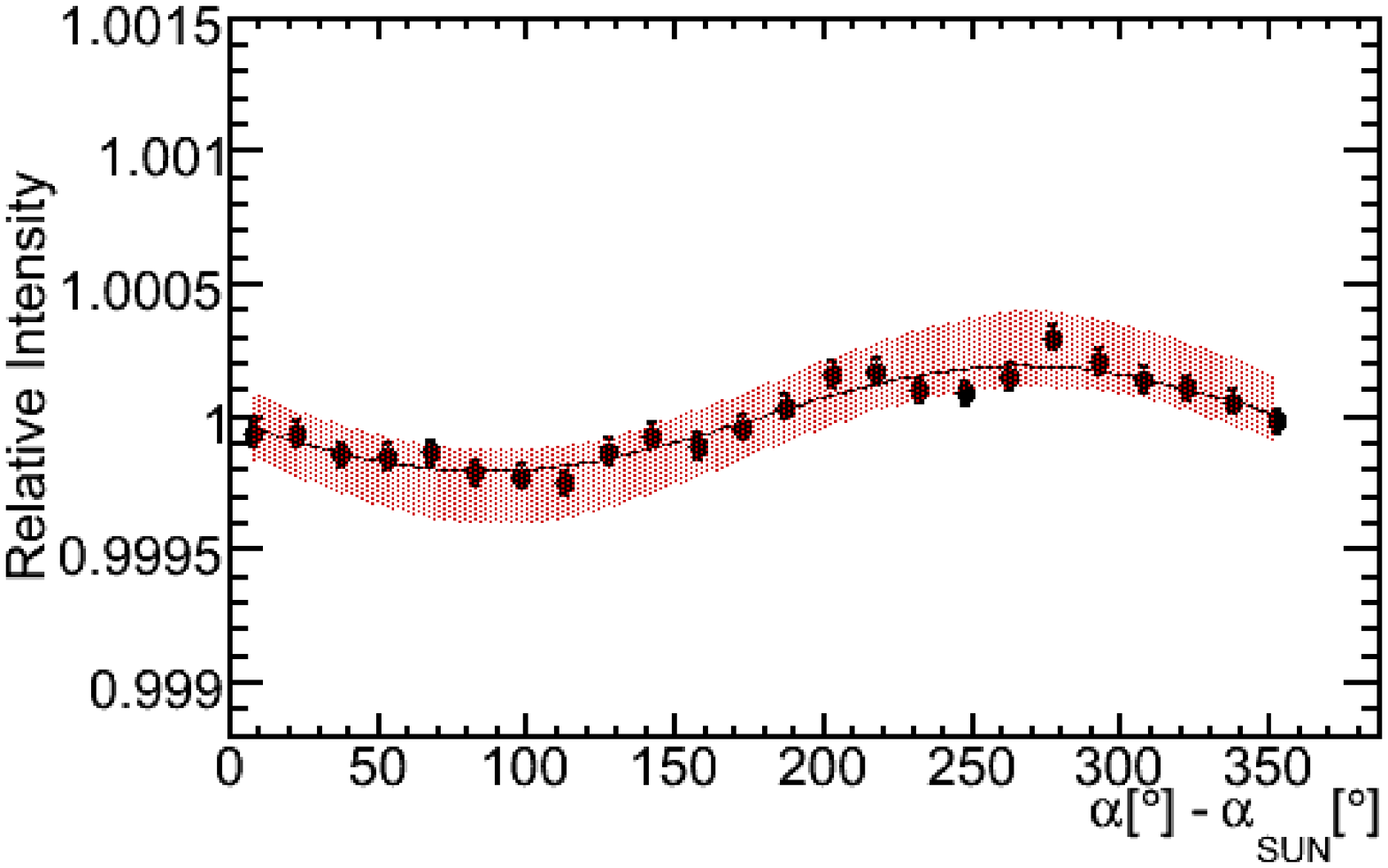}
}
\hspace{1cm}
\subfigure[] 
{
      \includegraphics[width=0.5\textwidth]{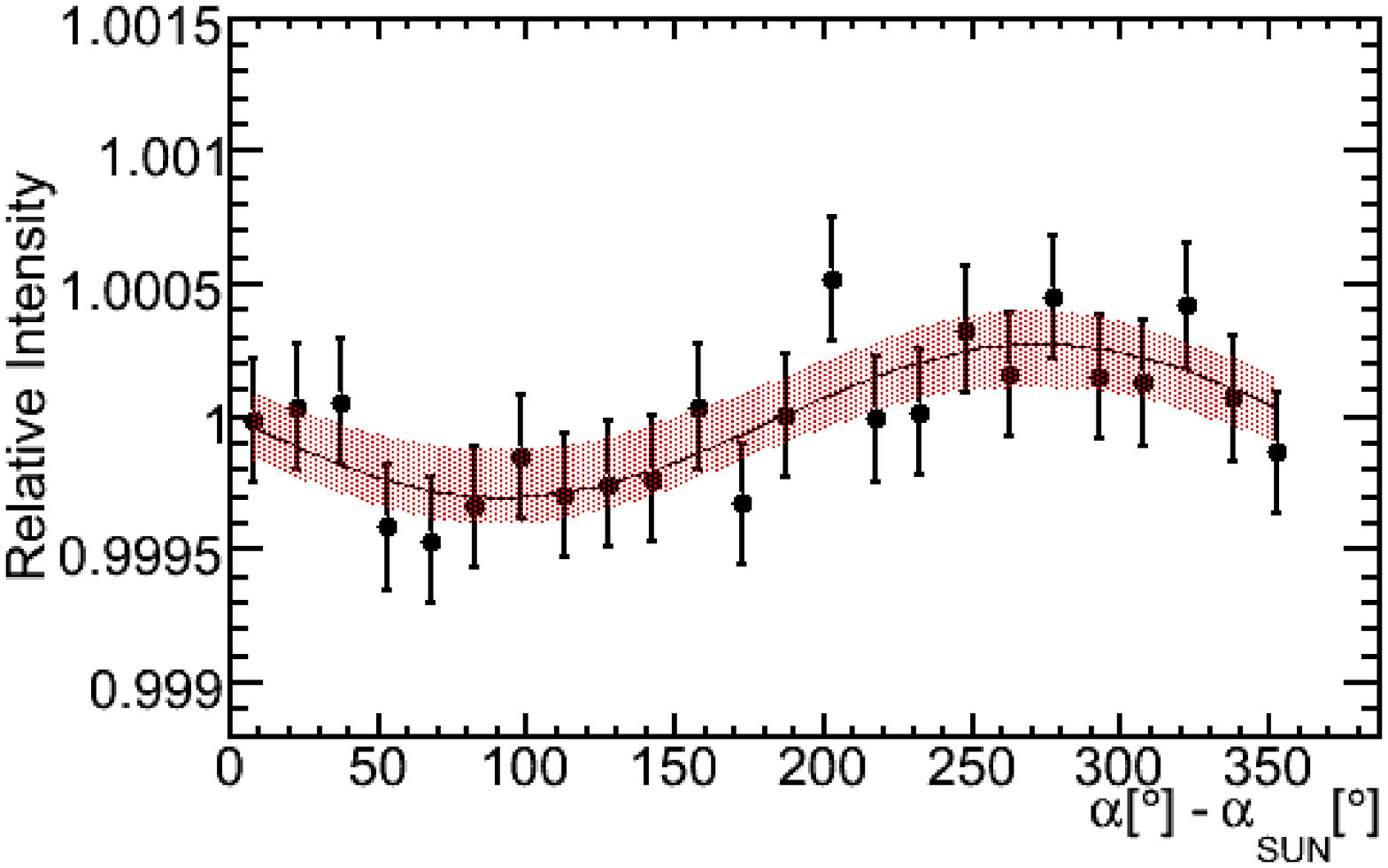}
}
\hspace{1cm}
\caption{\label{solskymap1d}
Figure (a) shows the one-dimensional projection in right ascension from the sun ($\alpha - \alpha_{sun}$) of the first energy band (20 TeV) of two-dimensional cosmic ray map in Figure~\ref{solskymap2d}-a. Figure (b) shows the  one-dimensional projection in right ascension from the sun ($\alpha - \alpha_{sun}$) of the second energy band (400 TeV) of two-dimensional cosmic ray map in Figure~\ref{solskymap2d}-b. The data are shown with statistical uncertainties, and the black line corresponds to the first and second harmonic fit to the data.}

\end{figure}

Figure~\ref{solskymap2d} and Figure~\ref{solskymap1d} show the maps of the cosmic ray arrival direction in solar reference frame, for both energy samples (20 and 400 TeV) along with with their projection onto right ascension relative to the Sun. The color scale is the relative intensity value for each pixel normalized to unity for each declination band. A fit to the projection of relative intensity distribution vs. ($\alpha - \alpha_{sun}$) was done using the first harmonic term of eq.~\ref{eq1}. Table~\ref{tabl:sol} shows the results of the first harmonic amplitude and phase along with $\chi^{2}/ndof$ of the fit.
\begin{table}[!ht]
\begin{ruledtabular}  
\caption{\label{tabl:sol} First harmonic fit values of the solar dipole anisotropy together with their statistical uncertainties for the energy bands centered at 20 TeV and 400 TeV.}
\begin{tabular*}{0.5\textwidth}{cccc}
\hline
$E_{Median}$ &  $A_{1_{SOL}}$ & $\phi_{1_{SOL}}$ & $\chi^{2}/ndof$  \\ 
(TeV)              & ($10^{-4}$)         & (degree)                    \\
\hline
20    &   $ 1.9 \pm0.1_{stat.}$ & $ 267.1 \pm 3.8_{stat.}$ & 23/21 \\ \hline
400  &   $ 2.9 \pm 0.7_{stat.}$ &$272.1 \pm 13.3_{stat.}$ & 12/21 \\ \hline

\end{tabular*}
\end{ruledtabular}   
\end{table}

To verify that the experimental observation of the solar dipole is consistent with expectation, the predicted projection of the solar anisotropy is calculated for the IceCube location. The expectation of the solar dipole was calculated by computing the relative intensity of the solar dipole through the cosmic ray plasma (eq.~\ref{eq:cg}). Instead of counting the number of events within a given bin in right ascension from the Sun, for each event, after time scrambling the data we calculated a mean weight corresponding to the expected relative intensity of the solar dipole .

The uncertainties in the cosmic ray spectral index, in the reconstructed arrival direction of the events, and the spread in the Earth's velocity over a year were included in the calculation of the uncertainty of the expectation. The mean spectral index was evaluated using the all-particle cosmic ray spectrum from~\cite{sim3} and the spectral index was found to be $\langle \gamma \rangle$ = 2.67$\pm$ 0.19. The value used for Earth's velocity was $v$ = 29.8$\pm$ 0.5 km/s~\citep{velocity}, where the error takes into account the spread between the maximum and the minimum along the elliptical orbit. The angle $\theta_{v}$ between the reconstructed direction of the muon events and the Earth's velocity vector at the time the event was detected was evaluated by accounting for the experimental point spread function. The expected solar dipole distribution, including the 68$\%$ spread in the uncertainty of the expectation, is shown as a shaded band in Figure~\ref{solskymap1d} for the low and high energy samples, respectively. The figures show that the observations are consistent with the expectation in both amplitude and phase for both low and high energy distribution. This demonstrates the reliability of the analysis to identify anisotropies at the level of a few $10^{-4}$, which supports the observations of the sidereal anisotropy.

\section{Systematic Uncertainties of the Sidereal Anisotropy}
\label{sec:syst}

In order to assess and quantify the systematic uncertainties in the sidereal anisotropy for the low and high energy samples of the cosmic ray arrival direction distribution, we performed two different studies, similar to~\citep{abbasi}. First of all, we estimated the statistical stability of the result and verified that the observation is unaffected by the particular choice of the data sample. Then we estimated the possible distortion effect on the sidereal anisotropy distribution derived from a possible annual modulation of the amplitude of the solar anisotropy.

\subsection{Data Stability}
\label{sec:datstab}
To assess the stability of the sidereal anisotropy, checks were applied by dividing the full data sample used in this analysis into series of two exclusive data sets by splitting both high and low energy data samples in halves based on different criteria. A full analysis was done with each dataset and the relative intensity distribution in right ascension was determined for each of them, along with a fit to the first and second harmonic term of eq.~\ref{eq1}, and compared to the ones from the complete low and high energy samples, respectively.

To check if the anisotropy had a seasonal dependence the data were divided into austral summer and austral winter sets. The summer set included events collected from December to May while the winter set included events collected from June to November. Since each dataset used in this test did not cover the full year, the sidereal anisotropy distribution was contaminated by the un-compensated solar dipole (see section~\ref{ssec:solar}). This spurious effect was accounted for by determining what the solar dipole should look like in a sidereal reference frame within the two seasonal time periods. In order to do so a numerical calculation was performed where, every 100$\mu$s, an event was generated with a unique UTC time, and with right ascension from the Sun sampled from the all-year experimental solar dipole distributions for each energy sample as shown in Figure~\ref{solskymap1d}. The corresponding distributions in the sidereal reference frame were then calculated and subtracted from the observed sidereal distribution in each seasonal time interval and the corrected sidereal distributions were then obtained.

To ensure that the sidereal anisotropy was not affected by uniform variations in rate, the daily median rate was determined and two data sets were selected. One dataset containing sub-runs with event rate above the median daily value, where a sub-run corresponded to approximately 2 minutes of observations, and one with event rate below the median daily rate. Once more the analysis was then applied to each dataset and the sidereal anisotropy distributions for these data sets were determined.

To check whether the measurement is stable against the choice of the particular event sample selection, two separate sub-run selection tests were applied. The first test was done by dividing the sub-runs randomly for each day in two halves, and the second by 
dividing in even- and odd-numbered sub-runs. The arrival direction distribution in sidereal reference frame was then determined for each of these data sets.

For each day good quality runs were selected that satisfied fundamental data integrity requirements. This run selection, along with sporadic data acquisition downtime resulted in data collection time gaps which represented about 10\% of the livetime in IceCube-59. To verify that the non-uniform time coverage due to gaps in the data was correctly handled by the azimuthal re-weighting procedure, a complete analysis was performed on the sub-sample of days with maximal data collection time (i.e. $\sim$24 hr). There were 214 such days during one calendar year of IceCube-59 physics run. The relative intensity was then determined for the cosmic ray arrival direction in sidereal reference frame for this data set.

\begin{figure}[!ht]
\includegraphics[width=0.5\textwidth]{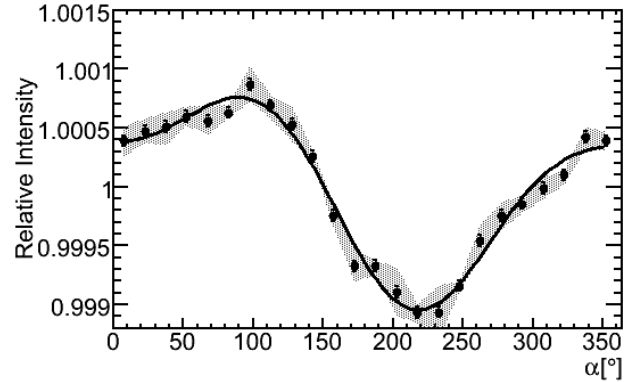}
\caption{\label{sysE1}
The one-dimensional projection in sidereal time frame of the two-dimensional cosmic ray map in Figure~\ref{sidskymap1d}-a for the 20 TeV band. The data are shown with statistical uncertainties, and the black line corresponds to the first and second harmonic fit to the data.
The gray band indicates the maximal spread from the stability checks.
}
\end{figure}

\begin{figure}[!ht]
\includegraphics[width=0.5\textwidth]{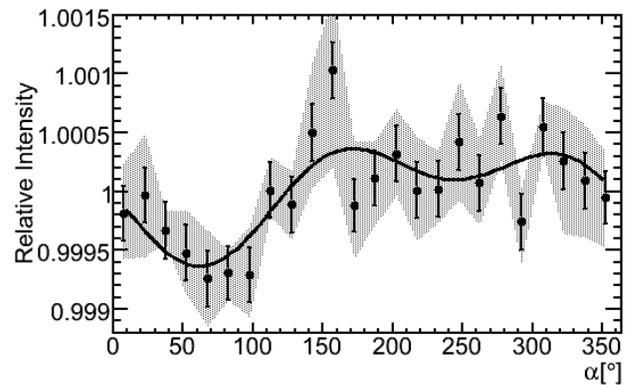}
\caption{\label{sysE2}
The one-dimensional projection in sidereal time frame of the two-dimensional cosmic ray map in Figure~\ref{sidskymap1d}-b for the 400 TeV band. The data are shown with statistical uncertainties, and the black line corresponds to a  the first and second harmonic fit to the data.
The gray band indicates the maximal spread from the stability checks.
}
\end{figure}

The sidereal distributions of relative intensity in the cosmic ray arrival direction for the low and high energy samples and for each of the above mentioned tests, were used to evaluate the spread in the experimental observation from the full-year event samples. The gray bands in Figure~\ref{sysE1} and~\ref{sysE2} describe the maximal spread obtained from the result of all the stability checks described in this section.

\subsection{anti-sidereal time}
\label{sec:anti}

The sidereal anisotropy will be distorted by the solar dipole unless data are collected within an integer number of full years. While the sidereal reference frame is defined where the celestial sky is fixed, the solar reference frame is defined where the Sun is fixed. This means that a sidereal day is on average 4 minutes shorter than a solar day, and therefore, while the solar time reference frame includes 365.25 days/year, the sidereal time reference frame  is composed of 366.25 days/year. A static point in the solar reference frame will move across the sidereal frame and return to the same position on the sky in one full year. As a consequence any static solar distribution averages to zero in sidereal reference frame after one year.

The situation however changes if for some reason the measured solar anisotropy has, for instance, an annual modulation of its amplitude. Since a non-static signal in solar reference frame does not average to zero in sidereal frame after one year, particular care is needed to account for this possible source of bias in the sidereal anisotropy. This introduces a bias in the reference frame where one day is 4 minutes shorter than a solar day (i.e. the sidereal frame) and an equivalent bias in the reference frame where one day is 4 minutes longer than a solar day. This defines the so-called anti-sidereal time, i.e. a non-physical reference frame obtained by reversing the sign of the transformation from solar time to sidereal time, where the anti-sidereal year consists of 364.25 days~\citep{nagashima83}. The antisidereal reference frame can, therefore, be used to quantify the distortion induced in the sidereal anisotropy~\citep{farley}.

\begin{figure}[!ht]
 \subfigure[] 
{
      \includegraphics[width=0.5\textwidth]{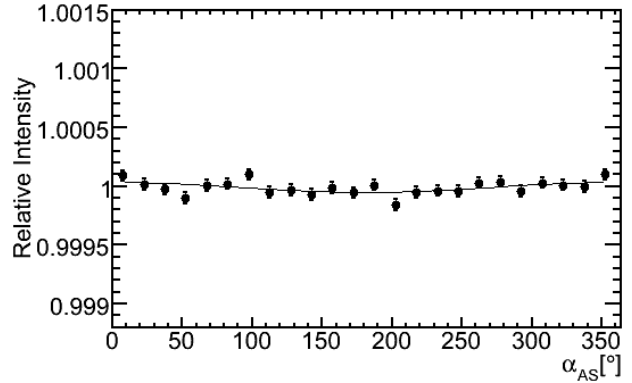}
}
\hspace{1cm}
\subfigure[] 
{
\includegraphics[width=0.5\textwidth]{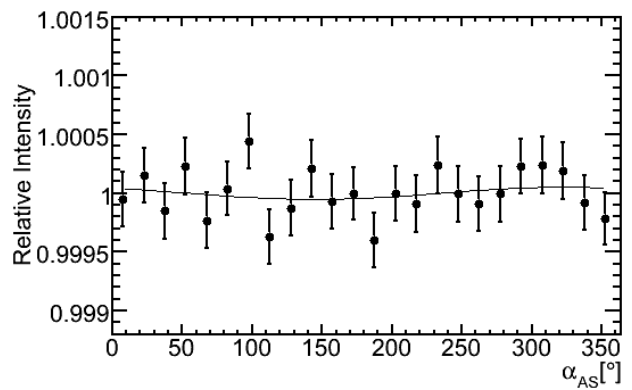}
}
\caption{\label{antisidSky}
Figure (a) shows the projection in $\alpha_{AS}$ of the relative intensity of cosmic ray arrival distribution using the anti-sidereal time for the low energy sample (median energy of the primary cosmic ray particle of 20 TeV).
Figure (b) shows the projection in $\alpha_{AS}$ of the relative intensity of cosmic rays arrival distribution for the high energy sample (median energy of the primary cosmic ray particle of 400 TeV). An anisotropy in the anti-sidereal reference frame is related to a distortion of the sidereal anisotropy induced by an annual modulation of the solar dipole amplitude.
}
\end{figure}

Figure~\ref{antisidSky} shows the relative intensity of cosmic rays arrival distribution in anti-sidereal reference frame (for the low and high energy samples). The anti-sidereal anisotropy is measured by using a coordinate system where the longitude coordinate is defined using the anti-sidereal time ($\alpha_{AS}$). The figure also shows a fit to the observed distributions with the dipole term of eq.~\ref{eq1} and Table~\ref{tabl:antisid} shows the fit results. Both the low and high energy samples show no significant observed amplitude in the anti-sidereal time. The uncertainty in the first harmonic amplitude and phase derived by the study of the anti-sidereal distribution was found to be within the statistical and systematic errors determined from the stability tests.

The results of all the systematic checks described in section~\ref{sec:datstab}, along with the estimate of the distortion in the sidereal anisotropy, based on the anti-sidereal distribution, were collectively used to estimate the global systematic uncertainties in the sidereal anisotropy fit parameters. Adding these systematic uncertainties the first and second harmonic amplitude and phase of the sidereal anisotropy for the low and high energy samples are summarized in Table~\ref{sidvals}.

\begin{ruledtabular}  
\begin{table}[!ht]
\caption{\label{tabl:antisid} First harmonic fit values of the anti-sidereal anisotropy for the energy bands centered at 20 TeV and 400 TeV.}

\begin{tabular*}{0.5\textwidth}{c cc cc cc}
\hline
$E_{Median}$ &  $A_{1_{ASID}}$ & $\phi_{1_{ASID}}$ & $\chi^{2}/ndof$ \\ 
(TeV)                & ($10^{-4}$)       & (degree)                 & \\
\hline
20    &    $ 0.4 \pm 0.1$  & $ 1.5 \pm 18.5$     & 29/21  \\ \hline
400  &    $ 0.5 \pm 0.7$  & $ 324.6 \pm 75.4$ &  17/21  \\ \hline
\hline
\end{tabular*}
\end{table}
\end{ruledtabular}   

\subsection{IceCube-40 String Sidereal Anisotropy}

In addition to the previously discussed systematic checks, an important cross-check is applied by looking at the result obtained from the previous year using the data collected from IceCube in its 40-strings configuration (IceCube-40) from May-2008 until May-2009. The same analysis described in this paper was applied to the IceCube-40 experimental data, along with the energy sample selection described in section~\ref{ssec:energy}. 

\begin{figure}[!ht]
\includegraphics[width=0.5\textwidth]{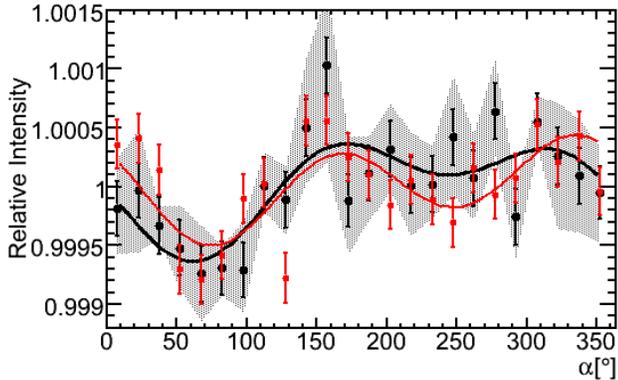}
\caption{\label{sidsysic59ic40}
This figure shows the IceCube-59 and IceCube-40  one-dimensional projections in sidereal time in black and red markers respectively at 400 TeV. The data are shown with statistical uncertainties for error bars. The line corresponds to the first and second harmonic fit to the data of IceCube-59 in black and IceCube-40 in red. The gray band indicates the estimated maximal systematic uncertainties of IceCube-59.
}
\end{figure}

The sidereal anisotropy observed at 20 TeV with IceCube-40 is found to be consistent with the reported observation with IceCube-22~\citep{abbasi} and with that observed using the IceCube-59 string configuration. Moreover, the relative intensity distribution for IceCube-40 as a function of right ascension for the 400 TeV band is also consistent with the distribution obtained with IceCube-59. Figure~\ref{sidsysic59ic40} shows the projection in right ascension of the relative intensity distribution at primary median energy of 400 TeV for both IceCube-40 (in red) and IceCube-59 (in black). The line corresponds to the first and second harmonic fit to the data of IceCube-59 in black and IceCube-40 in red. The gray band indicates the estimated maximal systematic uncertainties of IceCube-59. The results obtained with the two detector configurations are consistent within the statistical and systematic fluctuations. The stability of the result over different detector configuration supports the conclusion that the anisotropies observed at 20 and 400 TeV with IceCube-59 are real.

\section{Conclusions}
\label{sec:conc}

In this paper we presented the results on the large scale cosmic ray sidereal anisotropy, based on a total of 33$\times 10^{9}$ muon events collected by IceCube-59 from May 2009 to May 2010. In particular we showed the relative intensity in the arrival direction distribution at primary particle median energy of about 20 TeV and 400 TeV as shown in Figure~\ref{sidskymap2d}.

The relative intensity  distributions as a function of right ascension is fitted with a sum of first and second harmonic terms (eq.~\ref{eq1}). The amplitude and phase at 20 TeV and 400 TeV are summarized in section~\ref{sec:syst}. The observation of the sidereal anisotropy in the cosmic ray arrival direction is supported by the determination of the solar dipole expected from the Earth's revolution around the sun. The observed solar anisotropy agrees in amplitude and phase with the expectation in both energy bands. Moreover, the sidereal anisotropy is also supported by a number of data stability checks. One of these checks consisted of analyzing the data samples in the anti-sidereal time frame where no significant signal is observed. The observation of the solar dipole along with the absence of a signal in the anti-sidereal time frame in addition to all the stability tests, ensure the reliability of the sidereal anisotropy measurement for both 20 TeV and 400 TeV primary energy event samples.  

The sidereal anisotropy observed at 20 TeV with IceCube-59 is consistent with the previously reported observation with IceCube-22~\citep{abbasi}, thus providing a confirmation of a continuation of the arrival distribution pattern observed in the Northern equatorial hemisphere in the multi-TeV energy range. On the other hand the sidereal anisotropy observed at 400 TeV shows a significant relative deficit region in right ascension, $-6.3 \sigma$, where the excess is observed at median  primary energy of 20 TeV. In addition, the relative deficit region at low energy seems to have disappeared at median  primary energy of 400 TeV as shown in Figure~\ref{sigE1E2}. The observed anisotropy at 400 TeV shows substantial differences with respect to that observed at 20 TeV. Moreover, it does not show a continuation of the observations reported at high energies in the Northen hemisphere~\citep{amenomori},~\citep{aglietta}. This is the first significant anisotropy observed in cosmic ray arrival distribution in the 400 TeV range in the Southern hemisphere.

The sidereal anisotropy at 400 TeV also appears to be present in the data collected during the 40-string IceCube physics runs. The persistence of the anisotropy in IceCube-40 and IceCube-59 is an important verification that the anisotropies observed are not dependent on the detector configuration nor on the period the data were collected. Using events collected with the complete IceCube observatory (86-strings) will enable us to significantly improve the statistical power in the determination of sidereal anisotropy at a few hundreds TeV primary energy.

\section{Discussion}
\label{sec:disc}

The origin of the sidereal anisotropy is still unknown. It is believed that a possible contribution to this observed anisotropy might be from the Compton-Getting effect. The Compton-Getting dipole anisotropy we expect to see in this analysis is determined from Monte Carlo simulation and should appear with a maximum in the one-dimensional projection in right ascension between 290$^\circ$and 340$^\circ$ and a deficit between 110$^\circ$and 160$^\circ$ with an amplitude of $\sim 0.13\%$. In this model the cosmic rays are assumed to be at rest with respect to the Galactic center. The sidereal anisotropy from both energy samples do not appear to be consistent with expected Compton-Getting model~\citep{comptongetting} neither in amplitude nor in phase. However, it is possible that the Galactic cosmic ray rest frame has a smaller relative velocity and a different direction with respect to the one hypothesized in~\citep{comptongetting}. The cosmic ray rotation with respect to the Galactic center is complex and unknown, therefore, in this case we can only conclude that the cosmic rays are not at rest with respect to the Galactic center. 

It is also worth noting that when describing the Galactic cosmic ray propagation through diffusion models the large scale anisotropy is an important observable. The determination of cosmic ray anisotropy at median energy of 400 TeV could enable us to obtain an improved theoretical description of the diffusion processes of Galactic cosmic ray's energy ranges closer to the knee~\citep{berezinskii}.

We are continuously analyzing events from IceCube with updated configurations.  IceCube construction is now completed with 86 strings deployed with a volume of km$^{3}$ in January of 2011. With the higher statistical power expected from the observed cosmic ray muons we will be able to improve our understanding of the anisotropy and its energy dependence closer to the knee region. This will further our understanding of the propagation of cosmic rays and help to eventually reveal their sources.

\section{acknowledgments}

We acknowledge the support from the following agencies:
U.S. National Science Foundation-Office of Polar Programs,
U.S. National Science Foundation-Physics Division,
University of Wisconsin Alumni Research Foundation,
the Grid Laboratory Of Wisconsin (GLOW) grid infrastructure at the University of Wisconsin - Madison, the Open Science Grid (OSG) grid infrastructure;
U.S. Department of Energy, and National Energy Research Scientific Computing Center,
the Louisiana Optical Network Initiative (LONI) grid computing resources;
National Science and Engineering Research Council of Canada;
Swedish Research Council,
Swedish Polar Research Secretariat,
Swedish National Infrastructure for Computing (SNIC),
and Knut and Alice Wallenberg Foundation, Sweden;
German Ministry for Education and Research (BMBF),
Deutsche Forschungsgemeinschaft (DFG),
Research Department of Plasmas with Complex Interactions (Bochum), Germany;
Fund for Scientific Research (FNRS-FWO),
FWO Odysseus programme,
Flanders Institute to encourage scientific and technological research in industry (IWT),
Belgian Federal Science Policy Office (Belspo);
University of Oxford, United Kingdom;
Marsden Fund, New Zealand;
Japan Society for Promotion of Science (JSPS);
the Swiss National Science Foundation (SNSF), Switzerland;
A.~Gro{\ss} acknowledges support by the EU Marie Curie OIF Program;
J.~P.~Rodrigues acknowledges support by the Capes Foundation, Ministry of Education of Brazil.


\end{document}